\documentclass[12pt]{article}
\usepackage{amsfonts}

\newcommand{\sltr}{\ensuremath{S\!L(2,\mathbb{R})}}
\newcommand{\sltrmuo}{\ensuremath{S\!L(2,\mathbb{R})/U(1)}}

\newcommand{\be}{\begin{equation}}
\newcommand{\ee}{\end{equation}}
\newcommand{\ba}{\begin{eqnarray}}
\newcommand{\ea}{\end{eqnarray}}

\newcommand{\pd}{\partial}

\newcommand{\non}{\nonumber}

\newcommand{\G}{\mathcal{G}}     
\renewcommand{\H}{\mathcal{H}}   
\renewcommand{\S}{\mathcal{S}}   

\newcommand{\lc}{\epsilon}       

\renewcommand{\l}{\left}
\renewcommand{\r}{\right}
\newcommand{\h}{\frac{1}{2}}

\newcommand{\Tr}{{\rm Tr}}
 
\begin{document}

\title{\sc Thermal Correlators in Little String Theory}
\author{
    Philip A. DeBoer\thanks{pdeboer@physics.ubc.ca}~ and Moshe
    Rozali\thanks{rozali@physics.ubc.ca}\\ Department of Physics and
    Astronomy \\ University of British Columbia }
\date{\today}

\maketitle

We calculate, using holographic duality, the thermal two-point
function in finite temperature little string theory. The analysis of
those correlators reveals possible instabilities of the thermal
ensemble, as in previous discussions of the thermodynamics of little
string theory. We comment on the dependence of the instability on 
the spatial volume of the system.

\section{Introduction and Summary}

Little string theories (LST) \cite{BRS,seiberg} are strongly
interacting string theories, which exhibit phenomena quite different
from their critical counterparts. Most notably, they do not contain a
massless graviton in their spectrum. In their DLCQ definition
\cite{DLCQ} the strings appear in a very similar fashion to the
critical strings, with the important difference being that the
amplitude for splitting and joining strings is of order one, rather
than being a free parameter that can be taken to be arbitrarily small.

It is clearly of interest to study these theories and compare their
behavior to conventional field theories and to critical string
theories.  This allows for example an examination of nonlocal
theories, without having the complications associated with
gravity. Furthermore, this set of theories is very large, describing
generic singular limits of string theory compactifications.

Studies of the thermodynamics of LST, using holographic duality
\cite{holography}, reveal qualitatively different features from more
familiar theories \cite{us, kutasov,more}. The dual of LST at finite
temperature is the CHGS cigar geometry \cite{cghs}. In \cite{us} it was
pointed out that the infinite specific heat found in a tree level
approximation of the dual geometry is an artifact of examining the
latter in weak string coupling. The specific heat away from this limit
was argued later to be negative \cite{kutasov, kkk}. For a nice review
of these issues the reader may consult \cite{david}.

Negative specific heat is a generic feature of black holes in
asymptotically flat spaces, so perhaps one should not be too
surprised. However, the conjectured correspondence with a well-defined
boundary theory makes this feature somewhat mysterious. It would be
useful to find a clear interpretation of the implied instability of
the thermal ensemble. In \cite{kutasov} it was speculated that some
mode, found earlier in the context of zero temperature ``double
scaled'' LST \cite{gk}, could become tachyonic at one loop
level. Moreover, the analysis in \cite{kutasov} predicts an
interesting dependence of the instability on the spatial volume of the
system\footnote{Also there may be an interesting dependence on $k$,
the number of 5-branes. However, here and in \cite{kutasov} we are
working in leading order in the large $k$ limit.}.

However, the situation investigated in \cite{gk} is quite different ---
zero temperature theory far away on the Coulomb branch.  One of the
goals of the present paper is to formulate this conjecture, of being
on the verge of instability, in the context of LST at finite
temperature. Indeed we find that the system has modes which will
destabilize it, given the correct mass shift at one loop. Furthermore
we find complicated dependence of this phenomenon on the spatial
momenta of the modes, which is reminiscent of the claim in
\cite{kutasov}.

Specifically we study two point functions in the the thermal state of
LST. This is done using the holographic duals of these states, and
utilizing the results of \cite{teschner}.  This calculation has no
counterpart for a critical string theory, as Euclidean string theory
on a thermal circle has no on-shell vertex operators. Indeed, the
observables we concentrate on correspond to modes that are
non-normalizable in a radial direction, and therefore are naturally
interpreted as boundary rather than bulk properties. This is also the
reason LST has off-shell correlators which can be calculated
holographically, whereas critical string theory does not.

By and large, previous holographic probes of LST show patterns
consistent with a free theory, apart from some puzzles (cutoff on
off-shell momenta and some branch cuts in momentum space) which may
signal a breakdown of perturbation theory \cite{gk}.  Our correlators
show no free theory patterns. There are no stable states, or even
narrow resonances, revealed at that temperature. This is a sensible
behavior of a strongly interacting system at a finite temperature.

There are also some puzzles, similar to the ones in \cite{gk}, that
suggest the importance of small corrections at large enough
(Euclidean) frequencies. We comment on possible small modifications
of the results that would make the amplitudes sensible in the high
energy regime.  In particular, there are some subtleties in the
analytic continuation of the Euclidean results to real frequencies.
Therefore, it would be useful to perform a real time calculation, in
the gravity approximation, along the lines of \cite{real}. We hope to
return to that calculation in the future.

The organization of the paper is as follows. In section 2 we set up
the holographic dual to LST at finite temperature, review the relevant
world-sheet CFT, and construct the vertex operators.  In section 3 we
use that CFT to exhibit the thermal two-point function, using results
from \cite{teschner}, and discuss its properties.

\section{Vertex Operators}

\subsection{Review of \sltr\ Results}

We begin by constructing the vertex operators of \sltrmuo\ that
survive a finite temperature GSO projection. For the application to
LST, one needs only the non-normalizable modes in the holographic
background, since these correspond to observables in the dual
theory. For the vertex operators we concentrate on, such modes
originate from the discrete representations of \sltr\ and their
spectral flow images.

To begin, we need to set up the vertex operators of
\sltr. We use the notations of \cite{MO1,DVV}. The \sltr\ level $k$ WZW
action is
\be
    \S = \frac{k}{8\pi\alpha'}\int_{\pd\Sigma}
    d^2\sigma\Tr\l(g^{-1}\pd g~g^{-1}\pd g\r) + k\Gamma_{\rm WZ},
\ee
where $g$ is an element of \sltr\ and $\Gamma_{\rm WZ}$ is the
Wess-Zumino term
\be
    \Gamma_{\rm WZ} = \frac{1}{24\pi}\int_{\Sigma} d^3y \lc^{ijk}
    \Tr\l(g^{-1}\pd_ig~g^{-1}\pd_jg~g^{-1}\pd_kg\r).
\ee
The three-manifold $\Sigma$ has the world sheet as its boundary. Since
\sltr\ has no non-trivial 3-cycles, the level $k$ is not
quantized\footnote{In the full geometry $k$ appears as the level of
the WZW model on $\mathbb{S}^3$ and therefore is quantized.}.

The conserved currents of this action are $J^{3, \pm}_{L, R}$. As a
function of the left- and right-moving world sheet coordinates $z$ and
$\bar z$ the currents are given by
\be
    J^{3,\pm}_L(z) = k\Tr\l((T^{3,\pm})^*g^{-1}\pd g\r),
    \hspace{0.75cm} J^{3,\pm}_R(\bar z) = k\Tr\l(T^{3,\pm}\bar\pd
    g~g^{-1}\r),
\ee
where $T^{3,\pm}$ are the generators of \sltr. The currents can be
expanded,
\be \label{Jexpansion}
    J^{3,\pm}_L(z) = \sum^{\infty}_{n=-\infty} J^{3,\pm}_ne^{-inz},
    \hspace{0.75cm} J^{3,\pm}_R(\bar z) = \sum^{\infty}_{n=-\infty}
    \bar J^{3,\pm}_n e^{-in\bar z}.
\ee
The coefficients obey the Kac-Moody algebra:
\ba
    \l[J^3_n, J^3_m\r] & = & -\frac{k}{2}n\delta_{n, -m} \nonumber \\*
    \l[J^3_n, J^{\pm}_m\r] & = & \pm J^{\pm}_{n+m} \nonumber\\*
    \l[J^+_n, J^-_m\r] & = & -2J^3_{n+m} + k n \delta_{n, -m}.
\ea
The same algebra is satisfied by the modes of $\bar J$. The zero modes
$J_0^{3,\pm}$, $\bar J_0^{3,\pm}$ obey the $\sltr_L\times\sltr_R$
algebra, appropriate for a particle moving on the \sltr\ group
manifold.

The left-moving Virasoro generators $L_n$ are,
\be
    L_0 = \frac{1}{k-2}\l[\h(J^+_0J^-_0 + J^-_0J^+_0)-J^3_0J^3_0 +
    \sum_{m=1}^{\infty}(J^+_{-m}J^-_{m}+J^-_{-m}J^+_m-2J^3_{-m}J^3_{m})\r],
\ee
and for nonzero $n$
\be
    L_n = \frac{1}{k-2} \sum_{m=1}^{\infty}
    (J^+_{n-m}J^-_m+J^-_{n-m}J^+_m-2J^3_{n-m}J^3_m);
\ee
likewise for the right-moving generators $\bar L_n$. These obey the
Virasoro algebra with the central charge $c=\frac{3k}{k-2}$.

The unflowed representations are generated, as usual, on top of
representations of the zero mode algebra. We concentrate on the
discrete representations --- those are lowest (or highest) weight
representations of \sltr.

States in the discrete representations are described by two quantum
numbers $j$ and $m$. The quadratic Casimir operator is given by
$C_2(j)= \frac{1}{2}(J_0^+J_0^-+J_0^-J_0^+)-(J_0^3)^2 = -j(j-1)$ and
the eigenvalues of $J_0^3$ are given by $m$.  The quantum number $m$
takes on the values $j+n$ where $n$ is any nonnegative integer. The
state $|j;m=j\l.\r>$ is annihilated by $J_0^-$ as needed for a
lowest-weight state. Unflowed representations of the full Kac-Moody
algebra, based on these representations, are built in the usual way by
the action of creation operators $J_n^{3, \pm}$, $n<0$. These
representations are unitary, after imposing the Virasoro constraints,
for $\h < j<\frac{k-1}{2}$.

Additional representations of \sltr\ can be built using the operation
of spectral flow. Unlike in the $SU(2)$ case, the spectral flow
operation does not simply permute the representations, but rather
gives new representations. This operation takes $J^{3,\pm}_n \to
J^{3,\pm; w}_n$ given by
\be \label{Jflow}
    J^{3;w}_n = J^3_n - \frac{kw}{2}\delta_{n,0}, \hspace{0.75cm}
    J^{+;w}_n = J^+_{n+w}, \hspace{0.75cm} J^{-;w}_n = J^-_{n-w}.
\ee
The integer $w$ parameterizes the amount of spectral flow\footnote{We
consider here flow which acts the same way for both left and right
movers. The flow which acts with opposite signs will create winding
around the time-like direction of $\sltr$, when it is compact
\cite{arvind}.}.

Under spectral flow the Virasoro generators change as
\be
    L^w_n = L_n + wJ^3_n-\frac{kw^2}{4}\delta_{n,0},
\ee
and similarly for the $\bar L_n$.

The eigenvalues of $(J_0^w)^2$, $J_0^{3;w}$ are denoted by $-J(J-1)$ and
$M$ respectively. To maintain unitarity, after imposing the Virasoro
constraints, the eigenvalue $J$ of the flowed quadratic Casimir must
satisfy $\frac{1}{2}<J<\frac{k-1}{2}$.
 
We will concentrate on vertex operators which are tachyons in the
$\sltr/U(1)$ part of the geometry, namely states which are created
only by the zero modes of the $\sltr$ currents (with possible
oscillator modes from other parts of the CFT).  The vertex operators of
such states, in flowed discrete representations of \sltr\ with
eigenvalues $M$, ${\bar M}$, will be denoted $\Phi^w_{J M {\bar M}}$.

\subsection{Restriction to \sltrmuo}

To obtain the Euclidean cigar geometry the diagonal $U(1)$ subgroup of
the $\sltr_L\times\sltr_R$\ symmetry should be modded out.  This
$U(1)$ is generated by $ J = ( J^3_L + \bar{J^3_R} )$. The gauging is
described in detail in \cite{DVV}, and we follow their notation.

After introducing a gauge field and bosonizing, the action involves a
new compact scalar field, with the wrong-sign kinetic term,
\be
    \S_\phi = -(k/4\pi)\int d^2z\pd\phi\bar\pd\phi.
\ee
The original gauge symmetry is manifested by the requirement that all
physical operators must commute with the BRST charge
\be
    Q=\oint dz\;c(J^3_L + \frac{i}{2}k\pd\phi) + {\rm complex\ conjugate}.
\ee
The simplest vertex operators in the \sltrmuo\ are of the form
\be
    V = e^{ i q_L\phi_L(z) + i q_R \phi_R(\bar z) } \Phi^w_{J M \bar
    M}.
\ee
The BRST invariance then sets $q_L = -m$, $q_R = \bar{m}$. Note it is
the eigenvalues of the unflowed generators $J_0^3$, $\bar{J_0^3}$ which
appear here.

The compactification of $\phi$ further implies that
$\frac{q_L-q_R}{k}$ is an integer $n$, identified in \cite {DVV} as
the winding around the asymptotic circle of the cigar. This translates
to the restriction $\frac{M+\bar{M}}{k}$ is an integer $n-w$.  We will
also see that correlation functions depend only on this combination,
and not on $n$ and $w$ separately. We conclude therefore that without
loss of generality one can start with the unflowed representations of
\sltr\ --- the winding will be generated using the compact scalar
$\phi$ \footnote{This construction of vertex operators of $\sltr$, was
sketched in \cite{MO3}, using the language of parafermions, where the
last fact is more transparent.}.

The vertex operator has conformal weights (with respect to the
unflowed $L_0, \bar{L_0}$),
\ba
    h_{J p n } & = & -\frac{J(J-1)}{k-2} + \frac{(p + nk)^2}{4k},
    \nonumber\\ \bar h_{J p n } & = & -\frac{J(J-1)}{k-2} + \frac{(p -
    n k)^2}{4k},
\ea
where we denote $p=m-\bar{m}$ to be the momentum around the asymptotic
circle. This is the spectrum found in \cite{DVV}.  Note that it is the
Casimir operator with restricted domain which appears in the mass
formulas,
\ba
    \Delta = L_0 + \bar{L_0} - 1 &=& -\frac{J(J-1)}{k-2} +
    \frac{p^2}{2k}+ \frac{kn^2}{2}, \nonumber\\
         L_0 - \bar{L_0} &=& pn.
\ea
This justifies the interpretation of $p$, $n$ as momentum and winding,
respectively, around the Euclidean time direction.

We are interested in exploring correlators as a function of the
Euclidean momentum $p$. It is therefore essential not to restrict the
possible values of $p$. For that reason we choose $n=0$; otherwise the
level matching condition will make it difficult to construct a series
of allowed vertex operators with the different momenta, to be
interpreted as Fourier modes of a single object\footnote{Such vertex
operators will have to involve oscillator excitations which are
left-right asymmetric, and generally carry spacetime quantum numbers.}.
It may be expected that the set of observables does not change at
finite temperature, and the quantum number $n$ has no clear zero
temperature limit in the boundary theory.

However, we will see that the set of observables with $n=0$ does not
allow exploration of all possible momenta (for finite $k$). This is
essentially due to the same bound on the allowed off-shell momenta
found in \cite{gk}. In the present context, this puzzle may eventually
be resolved by making use of observables with $n \neq 0$.  We comment
further on this point below.

We therefore choose to work with graviton vertex operators, which in
addition to the wave function in the $\sltr/U(1)$ part have oscillator
excitations in the left and right moving sectors of the free CFT on
$\mathbb{R}^5$. The $\mathbb{S}^3$ part remains in its ground state.

\subsection{Worldsheet Supersymmetry}

To supersymmetrize the theory, eight Majorana-Weyl fermions are added
to each sector of the model. These fermions are free after a suitable
chiral rotation.

The addition of fermions changes the scaling dimension of vertex
operators. Since the zero point energy on the world-sheet is altered,
the term $\frac{2J(J-1)}{k(k-2)}$ is added to both $h_{Jpn}$ and $\bar
h_{Jpn}$ so that this scaling becomes
\be
    h_{J p n } = -\frac{J(J-1)}{k} + \frac{(p + nk)^2}{4k}
\ee
and likewise for $\bar h_{J p n }$.

The GSO projection used here breaks spacetime supersymmetry because it
imposes different periodicity conditions on spacetime bosons and
fermions around the compact Euclidean time direction. Spacetime
fermions are anti-periodic while bosons are periodic.

Requiring spacetime fermions to be anti-periodic, together with
modular invariance, introduces new phase factors into the sum over
spin structures. This was done by Atick and Witten \cite{AW} using the
functions $U_i(p, n)$, which modify the usual phases of Type II
strings at zero temperature. Explicitly\footnote{We are using here
the flat space result since we will be mainly interested in the
leading results in the large $k$ limit, and in the limit $k\rightarrow
\infty$ one recovers a flat space.}, these four functions are
\ba \non
    U_1(p,n) &=& \h\l(-1 + (-1)^p + (-1)^n + (-1)^{p+n}\r), \\ \non
    U_2(p,n) &=& \h\l( 1 - (-1)^p + (-1)^n + (-1)^{p+n}\r), \\ \non
    U_3(p,n) &=& \h\l( 1 + (-1)^p + (-1)^n - (-1)^{p+n}\r), \\
    U_4(p,n) &=& \h\l( 1 + (-1)^p - (-1)^n + (-1)^{p+n}\r).
\ea
Using the definition $U_i = U_i(p, n)$ for states with momentum $p$
and winding $n$ the desired one loop partition function is given by
the modular invariant trace
\ba \non
  && \Tr \l( e^{i\tau \H}
      \l[    \vphantom{(-1)^{\tilde F_{\sigma}}}
          U_3
            - (-1)^{F_{\tau}}U_4
            - (-1)^{F_{\sigma}}U_2
            - (-1)^{F_{\sigma}+F_{\tau}}U_1
        \r] \r. \\ \non
    && \hphantom{\Tr \l( e^{i\tau \H}\r.} \l.   
        \l[
          U_3
            - (-1)^{\tilde F_{\tau}}U_4
            - (-1)^{\tilde F_{\sigma}}U_2
            \pm (-1)^{\tilde F_{\sigma} + \tilde F_{\tau}}U_1
        \r]
    \r),
\ea
where $F_{\sigma, \tau}$ are the left-moving fermion number operators acting in
the $\sigma, \tau$ directions and $\H= L_0 + \bar{L_0}$ is the
Hamiltonian. The signs between the terms have been chosen to
correspond to the Type IIA,B string.

The vertex operators we are interested in survive the zero temperature
GSO projection.  The analysis in the previous  section confirms that all
momentum modes of them survive, enabling us to discuss the two point
function as an analytic function of frequency.

\subsection{Mass-Shell Conditions}

Each one of our vertex operator carries quantum numbers corresponding
to its dependence on various coordinates.  One of the general
properties of holography is the relation between on-shell bulk
quantities and off-shell boundary correlators. For this to be
possible, it is essential that the observables have a sufficient range
of bulk quantum numbers which are invisible from the boundary
perspective. One should be able to solve the bulk mass-shell
conditions for each value of the boundary quantum numbers.  In our
case the latter include the spatial momenta $q_i$, and the Matsubara
frequencies $p$. Even in the zero temperature analysis of
\cite{gk} this proves to be impossible, and we will see similar issues
arise here.

The graviton vertex operator in the full target space is given by
\be
    g_{ij}(q_i, p) = e^{-\phi-\bar\phi} \psi_{(i} \bar{\psi_{j)}}
    e^{iq_iX^i} \Phi_{J p },
\ee
where $\phi$, $\bar\phi$ are the bosonized superghost fields and $X^i$,
$\psi_i$, $\bar{\psi_i}$ are the free bosons and fermions in the
$\mathbb{R}^5$ component of the full CFT. The indices $i$, $j$ in this
expression are symmetrized. The vertex operator $\Phi_{Jp}$ is the
$\sltr/U(1)$ vertex operator discussed above, with the winding set to zero.
 
In writing $g_{ij}(q_i, p)$ as a function of $q_i$, $p$ only we imply
that other quantum numbers are to be regarded as implicit function of
those, given by the mass-shell condition. The mass-shell condition
arises as the requirement that $g_{ij}$ be a conformally invariant
vertex operator. Specifically,
\be
\label{mass}
     \frac{q^2}{2} - \frac{ J(J-1)}{k} + \frac{p^2}{4k} = 0,
\ee
so that $J$ is a function of the off-shell quantum numbers $q_i$,
$p$. In particular, the bound on $J$ translates here to a bound on
those quantum numbers, exactly as in \cite{gk}.

\section{Correlation Functions}

We are now ready to exhibit the thermal two-point functions of the
operators we are interested in. The ingredients of this calculation
were performed in \cite{teschner}, and reviewed in \cite{gk, MO3}.

In the notation of \cite{MO3} the two-point correlator of functions
(tachyon vertex operators) on $S\!L(2,\mathbb{C})/U(1)$, the Euclidean
version of
\sltr, can be calculated in the $x$-space (defined below) as
\be
    \l<\Phi_J(x_1)\Phi_J'(x_2)\r> = \frac{\delta(J-J')B(J)}
    {|z_{12}|^{2h}|\bar z_{12}|^{2\bar h}|x_{12}|^{4J}},
\ee
where $h,\bar{h}$ are the conformal weights, and $x_{12} = x_1-x_2$.
The $x$ basis is related to the $M$ basis, used in the present paper,
via
\be
    \Phi_{JM\bar M}=\int\frac{d^2x}{|x|^2}x^{J-M}x^{J-\bar
    M}\Phi_J(x).
\ee
This gives \cite{MO3}
\be
\label{tes}
    \l<\Phi_{JM\bar M}\Phi_{J'M'\bar M'}\r> =
    \frac{\pi\delta^2(M+M')\delta(J-J)B(J)} {|z_{12}|^{2h}|\bar
    z_{12}|^{2\bar h}\gamma(2J)} \frac{\Gamma(J+M)\Gamma(J-\bar M)}
    {\Gamma(1-J+M)\Gamma(1-J-\bar M)}.
\ee

The functions $B(J)$ and $\gamma(2J)$ are defined in
\cite{teschner,gk}:
\be
    B(J) = \frac{k\nu^{1-2J}}{\pi\gamma(\frac{2J-1}{k})},
\ee
and
\be
    \gamma(x) = \frac{\Gamma(x)}{\Gamma(1-x)}.
\ee
The constant $\nu$ and the function $B(J)$ both equal $1$ in the large
$k$ limit we are taking.

In calculating the spacetime two point function, one has to divide by
the volume of the conformal group. This normally makes the two point
function vanish, but in this context it is compensated by the volume
of target space \cite{seib, kutasov, MO3}. This leaves a finite piece,
which was calculated in \cite{MO3}.

In making the correlation function of our vertex operators $g_{ij}$,
the additional ingredients will make little difference. They will
modify the conformal weights in the denominator of (\ref{tes}), and
will multiply the correlator by a polarization dependent tensor in the
flat spatial directions.

The two point function of the vertex operators $g_{ij}(q_i, p)$ is
then given by:
\ba
 \l<g_{ij}(q_i, p)g_{kl}(q'_i, p')\r> \propto \,
 (\delta_{ik}\delta_{jl} + \delta_{ik}\delta_{jl}) \delta^5(q+q')\,
 \delta(p+p') \, \times \nonumber \\ \times \,
 \frac{1}{2J-1}\,\frac{1}{\gamma(2J)}\,\frac{\Gamma(J+M)\Gamma(J-\bar
 M)} {\Gamma(1-J+M)\Gamma(1-J-\bar M)}
\ea

Here $M, \bar{M}$ and $J$ are regarded as functions of the quantum
numbers $q_i$, $p$. We now proceed to comment on the structure of this
result and its analytic continuation to real frequencies.

\subsection{Euclidean Space}

The temperature Green's function can be written as a function of the
Matsubara frequencies $p$ as
\be
\label{corr}
  \G \sim \frac{1}{\gamma(1 + \sqrt{1+2kq^2+p^2})} \,
  \frac{\Gamma^2(\h + \h\sqrt{1+2kq^2+p^2}+\frac{p}{2})} {\Gamma^2(\h
  - \h\sqrt{1+2kq^2+p^2}+\frac{p}{2})}.
\ee
This is found using the conditions $M=p/2$, $\bar M=-p/2$, and the
solution
\be
    J = \h + \h\sqrt{1+2kq^2+p^2}
\ee
to the mass-shell condition (\ref{mass}).

A second solution to the mass-shell condition is discarded as it
violates the lower bound on $J$. The upper bound imposes a constraint
on a combination of the Euclidean energy $p$ and the transverse
momentum $q$,
\be
    \sqrt{1+2kq^2+p^2}<k-2.
\ee

There is an additional condition $M=J-n=p/2$, for $n$ an integer of
definite sign, since those are the allowed values in the discrete
representations. Positive $n$ corresponds to a highest-weight
representation, while negative $n$ corresponds to a lowest-weight
representation. This translates, using the mass-shell condition, to
\be
  \frac{kq^2}{2}+n^2 = J(2n-1).
\ee
Since $J$ is positive it is clear that $n$ must be positive as well,
so only the highest-weight representations are allowed.

So we see that only a discrete set of points in the frequency-momentum
space is obtained by the Euclidean string theory calculation. Our
approach is to analytically continue the result as needed, and attempt
interpretation of the correlator as a function of Lorentzian
frequencies\footnote{We note that the original derivation of the
correlators \cite{teschner} utilizes analytic continuation as well. It
would be interesting to explore alternative Lorentzian continuations,
along the lines of \cite{steve}.}.

\subsection{Minkowski Space}

The temperature Green's function $\G(p)$ can be analytically continued
to the retarded Green's function $G_R(\omega)$ using $p\to
-i\omega+\delta$. The analytic continuation of $\G$ is not unique
since it can, for example, be multiplied by a phase which vanishes at the
poles. Typically one requires a fall-off condition for large Euclidean
frequencies in order to fix the analytic continuation uniquely.

These fall-off conditions require knowledge of the correlators at
arbitrarily high Euclidean frequencies. As our frequencies are a
priori bounded (for any finite $k$), we are unable to discuss those
fall-off conditions. Even without modifying the correlator found by
Teschner (which is exact in tree level string theory), it is possible
that small modifications to the mass shell condition (\ref{mass}) will
result in satisfying the asymptotic fall-off conditions. We assume
this to be the case; as mentioned in the introduction a real time
calculation of the retarded Green's function will be useful to confirm
or refute this point.

We assume therefore that the form of the correlator (\ref{corr})
applies to Lorentzian frequencies as well.  The retarded Green's
function is therefore taken to be
\be
  G_R \sim \frac{\Gamma(-\sqrt{1+2kq^2-\omega^2}) \Gamma^2(\h +
  \h\sqrt{1+2kq^2-\omega^2}+\frac{i\omega}{2})} {\Gamma(1 +
  \sqrt{1+2kq^2-\omega^2}) \Gamma^2(\h -
  \h\sqrt{1+2kq^2-\omega^2}+\frac{i\omega}{2})}.
\ee

In a retarded Green's function an instability will manifest itself as a
pole (or other type of singularity, for more complex instabilities)
below the real axis. This will be a process which grows, rather than
decays, as a function of time. As we are looking for modes which may
destabilize the system upon including arbitrarily small corrections,
the interesting structure for our purposes occurs on the real
axis. This structure represents modes that are stable in the thermal
ensemble, an unusual feature.

The retarded Green's function found above has only one set of
singularities, coming from the factor
$\Gamma(-\sqrt{1+2kq^2-\omega^2})$. The singularities are located at
\be
  \omega = \pm\sqrt{1+2kq^2-n^2},
\ee
where $n$ is a nonnegative integer. These points are located along the
real and imaginary axes.  The points along the real axis are bounded
(for generic\footnote{There are some more possibilities at
special kinematic points, but as those will be sensitive to small
corrections, we concentrate on generic kinematics.} spatial momenta
$q_i$),
\be
    |\omega| < \sqrt{1+2kq^2}.
\ee

We see that a large number of possible instabilities is present for
non-zero spatial momenta $q^2$, a number of order ${k q^2}$. We note
that the holographic dual to LST contains a continuum of modes above a
gap, whose role in the theory is unclear. A typical momentum of those
modes, just above the gap, is $q^2 \sim \frac{1}{k}$. Therefore there
are order $k$ possible ``nearly destabilizing'' modes, and nearly
all of them are above the gap. The frequencies of those modes are much
smaller than $k$ (where our analysis is expected to receive corrections).

On the other hand, for zero spatial momentum there is only one such
possible mode, at $|\omega| = 1$\footnote{An additional potential
pole at $\omega=0$ is eliminated by a zero at that frequency.}. We see
a similar picture to the one suggested in \cite{kutasov}, where the
instability depends on the spatial volume (or the spatial momentum in
our case).

The retarded Green's function also has a set of zeros. These are on
the imaginary axis, so in particular they can eliminate a possible
divergence only for special values of $q_i$, the spatial momentum.

\section*{Acknowledgments}

We are thankful for useful conversations with Micha Berkooz, Mark
Laidlaw, Gordon Semenoff and Bill Unruh. The work was supported in
part by NSERC.


\begin{thebibliography}{99}

\bibitem{BRS}
M.~Berkooz, M.~Rozali and N.~Seiberg, ``Matrix Description of M-theory
on $T^4$ and $T^5$,'' Phys.\ Lett.\ B {\bf 408}, 105 (1997)
[arXiv:hep-th/9704089].
\bibitem{seiberg}
N.~Seiberg, ``New theories in six dimensions and matrix description of
M-theory on T**5 and T**5/Z(2),'' Phys.\ Lett.\ B {\bf 408}, 98 (1997)
[arXiv:hep-th/9705221].
\bibitem{DLCQ}
O.~Aharony, M.~Berkooz, S.~Kachru, N.~Seiberg and E.~Silverstein,
``Matrix description of interacting theories in six dimensions,''
Adv.\ Theor.\ Math.\ Phys.\ {\bf 1}, 148 (1998)
[arXiv:hep-th/9707079];

E.~Witten, ``On the conformal field theory of the Higgs branch,'' JHEP
{\bf 9707}, 003 (1997) [arXiv:hep-th/9707093].
\bibitem{holography}
O.~Aharony, M.~Berkooz, D.~Kutasov and N.~Seiberg, ``Linear dilatons,
NS5-branes and holography,'' JHEP {\bf 9810}, 004 (1998)
[arXiv:hep-th/9808149].
\bibitem{us}
T.~Harmark and N.~A.~Obers, ``Hagedorn behaviour of little string
theory from string corrections to NS5-branes,'' Phys.\ Lett.\ B {\bf
485}, 285 (2000) [arXiv:hep-th/0005021].

M.~Berkooz and M.~Rozali, ``Near Hagedorn dynamics of NS fivebranes,
or a new universality class of coiled strings,'' JHEP {\bf 0005}, 040
(2000) [arXiv:hep-th/0005047].
\bibitem{kutasov}
D.~Kutasov and D.~A.~Sahakyan, ``Comments on the thermodynamics of
little string theory,'' JHEP {\bf 0102}, 021 (2001)
[arXiv:hep-th/0012258].
\bibitem{more}
K.~Narayan and M.~Rangamani, ``Hot little string correlators: A view
from supergravity,'' JHEP {\bf 0108}, 054 (2001)
[arXiv:hep-th/0107111].

A.~Buchel, ``On the thermodynamic instability of LST,''
[arXiv:hep-th/0107102].

M.~Rangamani, ``Little string thermodynamics,'' JHEP {\bf 0106}, 042
(2001) [arXiv:hep-th/0104125].

T.~Harmark and N.~A.~Obers, ``Hagedorn behavior of little string
theories,'' [arXiv:hep-th/0010169].
\bibitem{cghs}
C.~G.~Callan, S.~B.~Giddings, J.~A.~Harvey and A.~Strominger,
``Evanescent Black Holes,'' Phys.\ Rev.\ D {\bf 45}, 1005 (1992)
[arXiv:hep-th/9111056].

E.~Witten, ``On string theory and black holes,'' Phys.\ Rev.\ D {\bf
44}, 314 (1991).
\bibitem{kkk}
V.~Kazakov, I.~K.~Kostov and D.~Kutasov, ``A matrix model for the
two-dimensional black hole,'' Nucl.\ Phys.\ B {\bf 622}, 141 (2002)
[arXiv:hep-th/0101011].
\bibitem{david}
D.~Kutasov, ``Introduction to little string theory,''
{\it Prepared for ICTP Spring School on Superstrings and Related
Matters, Trieste, Italy, 2-10 Apr 2001}.
\bibitem{gk}
A.~Giveon and D.~Kutasov, ``Little string theory in a double scaling
limit,'' JHEP {\bf 9910}, 034 (1999) [arXiv:hep-th/9909110].

A.~Giveon and D.~Kutasov, ``Comments on double scaled little string
theory,'' JHEP {\bf 0001}, 023 (2000) [arXiv:hep-th/9911039].
\bibitem{teschner}
J.~Teschner, ``On structure constants and fusion rules in the
SL(2,C)/SU(2) WZNW model,'' Nucl.\ Phys.\ B {\bf 546}, 390 (1999)
[arXiv:hep-th/9712256].

J.~Teschner, ``The mini-superspace limit of the SL(2,C)/SU(2) WZNW
model,'' Nucl.\ Phys.\ B {\bf 546}, 369 (1999) [arXiv:hep-th/9712258];

J.~Teschner, ``Operator product expansion and factorization in the
H-3+ WZNW model,'' Nucl.\ Phys.\ B {\bf 571}, 555 (2000)
[arXiv:hep-th/9906215].
\bibitem{steve}
P.~Kraus, H.~Ooguri and S.~Shenker,
``Inside the horizon with AdS/CFT,''
[arXiv:hep-th/0212277].
\bibitem{real}
C.~P.~Herzog and D.~T.~Son, ``Schwinger-Keldysh propagators from
AdS/CFT correspondence,'' [arXiv:hep-th/0212072];

G.~Policastro, D.~T.~Son and A.~O.~Starinets, ``From AdS/CFT
correspondence to hydrodynamics,'' JHEP {\bf 0209}, 043 (2002)
[arXiv:hep-th/0205052].

G.~Policastro, D.~T.~Son and A.~O.~Starinets, ``From AdS/CFT
correspondence to hydrodynamics. II: Sound waves,''
[arXiv:hep-th/0210220].

C.~P.~Herzog, ``The hydrodynamics of M-theory,'' JHEP {\bf 0212}, 026
(2002) [arXiv:hep-th/0210126].

D.~T.~Son and A.~O.~Starinets, ``Minkowski-space correlators in
AdS/CFT correspondence: Recipe and applications,'' JHEP {\bf 0209},
042 (2002) [arXiv:hep-th/0205051].
\bibitem{MO1}
J.~M.~Maldacena and H.~Ooguri, ``Strings in AdS(3) and SL(2,R) WZW
model. I,'' J.\ Math.\ Phys.\ {\bf 42}, 2929 (2001)
[arXiv:hep-th/0001053].

J.~M.~Maldacena, H.~Ooguri and J.~Son, ``Strings in AdS(3) and the
SL(2,R) WZW model. II: Euclidean black hole,'' J.\ Math.\ Phys.\ {\bf
42}, 2961 (2001) [arXiv:hep-th/0005183].
\bibitem{DVV}
R.~Dijkgraaf, H.~Verlinde and E.~Verlinde, ``String propagation in a
black hole geometry,'' Nucl.\ Phys.\ B {\bf 371}, 269 (1992).
\bibitem{arvind}
A.~Rajaraman and M.~Rozali, ``Boundary states for D-branes in
AdS(3),'' Phys.\ Rev.\ D {\bf 66}, 026006 (2002)
[arXiv:hep-th/0108001].
\bibitem{MO3}
J.~M.~Maldacena and H.~Ooguri, ``Strings in AdS(3) and the SL(2,R) WZW
model. III: Correlation functions,'' Phys.\ Rev.\ D {\bf 65}, 106006
(2002) [arXiv:hep-th/0111180].
\bibitem{AW}
J.~J.~Atick and E.~Witten, ``The Hagedorn Transition And The Number Of
Degrees Of Freedom Of String Theory,'' Nucl.\ Phys.\ B {\bf 310}, 291
(1988).
\bibitem{seib}
N.~Seiberg, ``Notes On Quantum Liouville Theory And Quantum Gravity,''
Prog.\ Theor.\ Phys.\ Suppl.\ {\bf 102}, 319 (1990).

\end{thebibliography}
\end{document}